\documentclass[apj]{emulateapj}
\usepackage{mathptmx}

\def\gtorder{\mathrel{\raise.3ex\hbox{$>$}\mkern-14mu
             \lower0.6ex\hbox{$\sim$}}}
\def\ltorder{\mathrel{\raise.3ex\hbox{$<$}\mkern-14mu
             \lower0.6ex\hbox{$\sim$}}}

\slugcomment{Draft of \today}

\shorttitle{Spectroscopic Redshifts for Seven Lens Galaxies}

\shortauthors{Ofek et al.}

\begin{document}

\title{Spectroscopic Redshifts for Seven Lens Galaxies}
\author{
Eran~O.~Ofek,\altaffilmark{1}
Dan~Maoz,\altaffilmark{1}
Hans-Walter~Rix,\altaffilmark{2}
Christopher~S. Kochanek,\altaffilmark{3} and
Emilio~E.~Falco\altaffilmark{4}
}
\altaffiltext{1}{School of Physics and Astronomy and the Wise Observatory, Tel-Aviv University, Tel-Aviv 69978, Israel. Electronic address: (eran, dani)@wise.tau.ac.il}
\altaffiltext{2}{Max-Planck-Institut f\"{u}r Astronomie, K\"{o}nigstuhl 17, D-69117 Heidelberg, Germany}
\altaffiltext{3}{Department of Astronomy, The Ohio State University, 140 West 18th Avenue, Columbus, OH 43210}
\altaffiltext{4}{Harvard-Smithsonian Center for Astrophysics, 60 Garden Street, Cambridge, MA 02138}

\begin{abstract}

%
We report Very Large Telescope observations of $11$ lensed quasars,
designed to measure the redshifts of their lens galaxies.
We successfully determined the redshifts for seven systems,
five of which were previously unknown.
The securely measured redshifts for the lensing galaxies are:
HE0047$-$1756 $z=0.408$;
PMNJ0134$-$0931 $z=0.766$;
HE0230$-$2130 $z=0.522$;
HE0435$-$1223 $z=0.455$;
SDSS0924$+$021 $z=0.393$; 
LBQS1009$-$025 $z=0.871$; and
WFIJ2033$-$472 $z=0.658$.
For four additional systems
(BRI0952$-$0115, Q1017$-$207, Q1355$-$2257 and PMNJ1632$-$003)
we estimate tentative
redshifts based on some features in their spectra.

\end{abstract}


\section{Introduction}
\label{Introduction}


The roughly $90$ known gravitational lens 
systems are powerful tools for studying cosmology,
galaxy structure and galaxy evolution.
Most systems have been observed with the
{\it Hubble Space Telescope} (HST) to obtain
precise photometry and astrometry, and increasing numbers have
time delay measurements, measurements of microlensing variability,
and determinations of the velocity dispersion of the lens galaxy.
However, about $60\%$ of the systems still lack the most basic parameters
needed to use a lens as an astrophysical tool --
the lens and/or source redshifts.


One important application of the lenses is to study the evolution of galaxy
mass-to-light ratios.  Gravitational lenses, 
unlike other samples of galaxies,
are selected based on their masses, rather than their colors 
or surface brightnesses. By combining the masses 
of the lenses, as determined by lensing, with the photometry
of the lens galaxies as determined from HST images,
it is possible to measure directly
the mass-to-light ratio of the lens population as a function of redshift
(e.g., Treu et al. 2002; Rusin et al. 2003; Treu \& Koopmans 2004;
Rusin \& Kochanek 2005).
Rusin \& Kochanek (2005), found that  mass-selected lens
galaxies show the same evolution rate as has been found for early-type 
galaxies in clusters (e.g., van Dokkum et al 1998;
Pahre et al. 2001),
and agree with the field evolution rates found
by van Dokkum et al. (2001), but not with the faster rates
found by Treu et al. (2002).
The sample studied by the latter authors may have selection effects biasing
it towards higher surface brightness, younger, systems.
Many of the lens mass-to-light ratios have, however, large
errors due to the uncertainties in their
redshifts, increasing the errors in the estimated average evolution rate.

A second important application of lenses is to determine the surface
density of dark matter and stars near the lensed images, by measuring
time delays between lensed images, and by monitoring for microlensing. 
Time delays determine
the combination of physical variables 
$\Delta t \propto H_0 (1-\langle\kappa\rangle)$
(Kochanek 2002),
where $H_0$ is the Hubble parameter,
$\langle\kappa\rangle=\Sigma/\Sigma_c$ is the mean surface 
density in the annulus between the images used to determine
the time delay, normalized to the critical surface density,
\begin{equation}  
\Sigma_{c}=\frac{c^{2}}{4\pi G}\frac{D_{s}}{D_{l} D_{ls}}.
\end{equation}
The critical surface density is a function of the
angular diameter distances
$D_{s}$, $D_{l}$, and $D_{ls}$,
between the observer and
source, observer and lens, and lens and source, respectively.
Assuming $H_{0}$ is known, measurements of time delays and
microlensing variability can be used to determine the surface mass
density and the fraction of that density in stars
in the interesting regime where a galaxy is changing
from being primarily stars to being dark-matter dominated.
%
%

A third application of lensed quasar systems which requires the
lens galaxy redshifts is the 
estimation of the mass and number evolution of massive galaxies.
For an unevolving comoving
density of massive galaxies, the number of lenses is proportional to the
volume out to the source, which is determined by cosmology 
(e.g., Mitchell et al. 2004).
Thus, no-evolution 
models with large dark-energy content, which have a large volume
out to a given source redshift, predict a large incidence of quasar lensing
and a distribution of lens galaxy redshifts shifted to relatively high redshift.
The currently favored cosmology
($\Omega_m\simeq 0.3$, $\Omega_{\Lambda}\simeq 0.7$) may predict too 
many lenses, given a non-evolving lens population (e.g., Maoz 2005).
We can therefore turn the problem around; for a fixed cosmology,
the only significant variables are the mass and number-density of
massive galaxies, and their evolution
(e.g., Ofek, Rix, \& Maoz 2003; Chae et al. 2003).
By analyzing samples having complete source and lens redshifts, 
it is possible to constrain the evolution rate of 
a mass-selected sample, without the problem of surface 
brightness selection effects that plague other samples.

In summary, each new lens redshift facilitates a wide array of astrophysical
experiments.
%
In this paper, we present European Southern Observatory (ESO) Very Large Telescope (VLT)
Focal Reducer/Low Dispersion Spectrograph - 2 (FORS2)
observations of $11$ lensed systems,
most of which lack lens galaxy redshift information.
In seven cases the spectra yield reliable redshifts for the lensing
galaxies.
In \S\ref{Observations} we describe the observations, the
reduction, and the isolation and characterization
of the lens galaxy spectra in the presence of
strong contamination by lensed quasar light.
The spectra
and the measured redshifts for each galaxy
are presented and discussed in \S\ref{Notes}.

\section{Observations and Reduction}
\label{Observations}

We obtained low resolution spectra for $11$ gravitationally lensed systems
using the FORS2 spectrograph mounted on the VLT-Unit Telescope $1$
at Paranal.
We used 
a $0\farcs7$-wide slit and
the 200I grism blazed at $7450$~\AA,
giving a wavelength range of $5500$~\AA~to $9000$~\AA,
a dispersion of $2.5$~\AA~per binned pixel, and
$\sim5$~\AA~full width at half maximum (FWHM) resolution.
The MIT $2$k$\times4$k pixel CCD was used, with on-chip binning of
$2\times2$, giving a scale of $0\farcs252$~per binned pixel.
The observations were conducted in Service Mode
on various dates,
with an image quality varying
between $0\farcs5$ and $0\farcs9$ FWHM.
Table~\ref{Table-Obs} lists, for each object,
the observation date, the number of
exposures and integration time per night, the
total number of exposures,
the total integration time, 
the seeing range and median in which the images were obtained,
and the slit position angle.

\begin{figure*}
\centerline{\includegraphics[width=18.0cm]{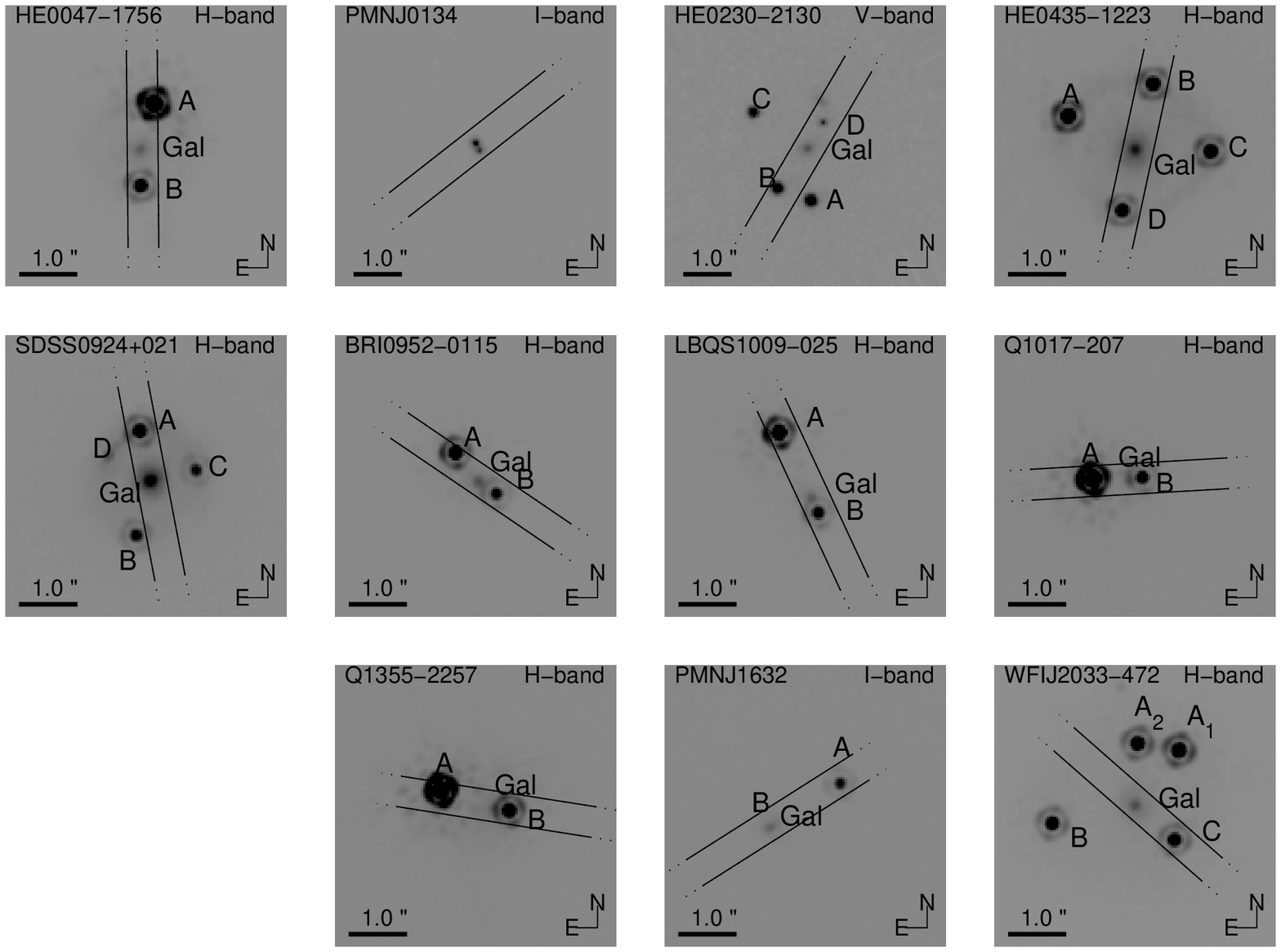}}
\caption {HST images of the lensed systems. The
VLT/FORS2 slit orientations, and actual slit widths are overlayed.
\label{AllFindingCharts} }
\end{figure*}

\begin{deluxetable*}{lrrccr}
\tablecolumns{6}
\tablewidth{0pt}
\tablecaption{Log of observations}
\tablehead{
\colhead{Object Name} &
\colhead{\# of Exp.} &
\colhead{Exp. time} &
\colhead{Date} &
\colhead{Seeing (Min..Median..Max)} &
\colhead{PA} \\
\colhead{} &
\colhead{} &
\colhead{[sec]} &
\colhead{} &
\colhead{[``]} &
\colhead{[deg]}
}
\startdata
HE0047$-$1756   &  $1$ &  $1800$ & 15-10-2004 &                                        &          \\
                &  $1$ &  $1800$ & 13-11-2004 &                                        &          \\
{\it total}     &  $2$ &  $3600$ &            & $0\farcs64$..$0\farcs73$..$0\farcs82$  &   $0.47$ \\
\hline
PMNJ0134$-$0931 &  $1$ &  $1200$ & 13-11-2004 & $0\farcs79$..$0\farcs79$..$0\farcs79$  & $-52.52$ \\
\hline
HE0230$-$2130   &  $1$ &  $1800$ & 11-11-2004 & $0\farcs87$..$0\farcs87$..$0\farcs87$  & $-30.38$ \\
\hline
HE0435$-$1223   &  $1$ &   $900$ & 14-12-2004 & $0\farcs75$..$0\farcs75$..$0\farcs75$  & $-12.20$ \\
\hline
SDSS0924$+$021  &  $2$ &  $2900$ & 14-12-2004 &                                        &          \\
                &  $2$ &  $2900$ & 15-01-2005 &                                        &          \\
{\it total}     &  $4$ &  $5800$ &            & $0\farcs62$..$0\farcs71$..$1\farcs39$  &  $10.83$ \\
\hline
BRI0952$-$0115  &  $6$ &  $6350$ & 05-01-2005 &                                        &          \\
                &  $4$ &  $5800$ & 14-01-2005 &                                        &          \\
{\it total}     & $10$ & $12150$ &            & $0\farcs63$..$0\farcs77$..$1\farcs03$  &  $55.78$ \\
\hline
LBQS1009$-$025  &  $2$ &  $2900$ & 08-01-2005 &                                        &          \\
                &  $4$ &  $5800$ & 14-01-2005 &                                        &          \\
{\it total}     &  $6$ &  $8700$ &            & $0\farcs68$..$0\farcs78$..$1\farcs05$  &  $24.95$ \\
\hline
Q1017$-$207     &  $2$ &  $2900$ & 12-01-2005 &                                        &          \\
                &  $2$ &  $2900$ & 15-01-2005 &                                        &          \\
{\it total}     &  $4$ &  $5800$ &            & $0\farcs65$..$0\farcs79$..$0\farcs90$  & $-86.65$ \\
\hline
Q1355$-$2257    &  $3$ &  $4350$ & 04-02-2005 &                                        &          \\
                &  $2$ &  $2900$ & 14-03-2005 &                                        &          \\
{\it total}     &  $5$ &  $7250$ &            & $0\farcs71$..$1\farcs01$..$1\farcs26$  & $81.03$  \\
\hline
PMNJ1632$-$003  &  $2$ &  $2900$ & 14-03-2005 & $0\farcs57$..$0\farcs63$..$0\farcs68$  & $-57.60$ \\
\hline
WFIJ2033$-$472  &  $1$ &  $1800$ & 15-10-2004 & $0\farcs71$..$0\farcs71$..$0\farcs71$  & $48.72$  \\
\hline
\enddata
\tablecomments{For objects that were observed on more then one night,
the last line gives the total number and total time of exposures.}
\label{Table-Obs}
\end{deluxetable*}
Figure~\ref{AllFindingCharts} shows
HST images of each system,
with the slit orientation and actual slit width overlayed.
Even with the good seeing conditions at the VLT site, 
the lensing galaxy light is blended with light from
one or more of the quasar images.
The slit orientation was chosen to
include, in addition to the lens galaxy,
the lensed quasar image producing the dominant
contamination.
With a proper characterization of the contaminating image,
we could then subtract it from the lens spectrum.

Image reduction was performed
on the bias-subtracted and wavelength-calibrated
images provided by ESO
using MATLAB scripts specifically written by us for this purpose.
The process included the following steps.
First, cosmic-ray rejection was performed by interpolating over pixels
with counts
that were $10\sigma$ above their neighboring pixels.
We then improved the wavelength calibration by identifying the night-sky emission line peaks,
and fitting a third order polynomial matching the measured wavelength of the peaks
to their known wavelengths\footnote{Sky line wavelengths were taken from\\ http://alamoana.keck.hawaii.edu/inst/lris/skylines.html}.
Sky subtraction was performed by means of a linear fit to the background counts
in each CCD row along the cross-dispersion direction.
In the fit, we used a $3.5\sigma$-clipping algorithm to remove outlier points due to residual
cosmic rays and bad pixels.

In all cases, the galaxy light is blended with the light from the quasar images.
Due to the different colors of the components
and the faintness of the galaxy relative to the quasar,
conventional tracing of the positions of the spectra along
the cross-dispersion direction may not work.
In order to overcome this problem,
we have designed our tracing algorithm as follows.

\begin{figure}
\centerline{\includegraphics[width=8.5cm]{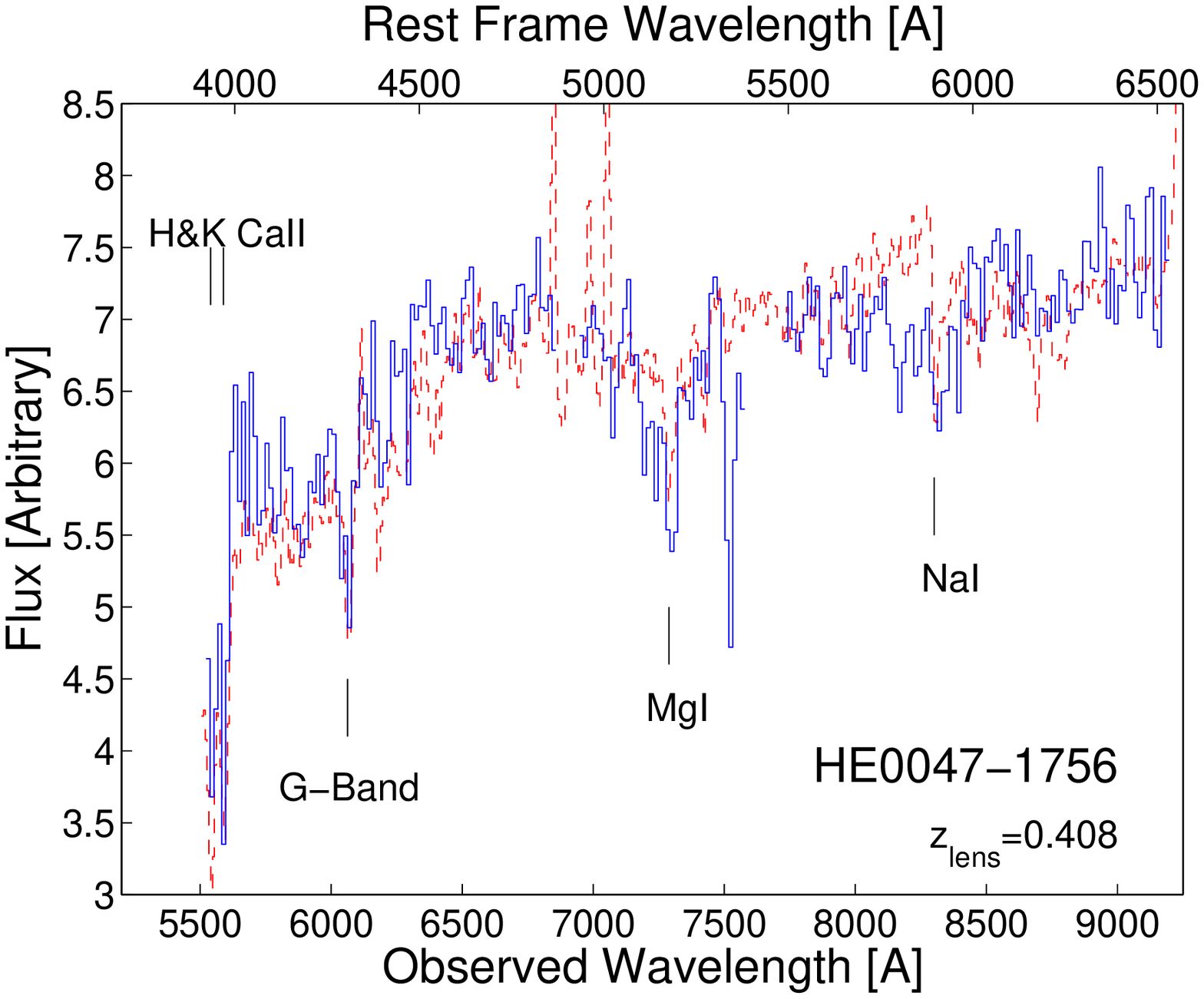}}
\caption {VLT/FORS2 spectrum, in the observer frame, of the lensing galaxy (solid line) of
HE0047$-$1756, after subtraction of the contaminating
quasar light. The dashed line shows the best fit galaxy template,
after redshifting and correcting for
Galactic extinction (Schlegel et al. 1998).
The main absorption features are marked.
Sections of the spectrum with strong telluric absorptions
($6860$-$6930$~\AA~and $7580$-$7720$~\AA)~
have been excised from the spectrum.
\label{HE0047} }
\end{figure}

\begin{figure}
\centerline{\includegraphics[width=8.5cm]{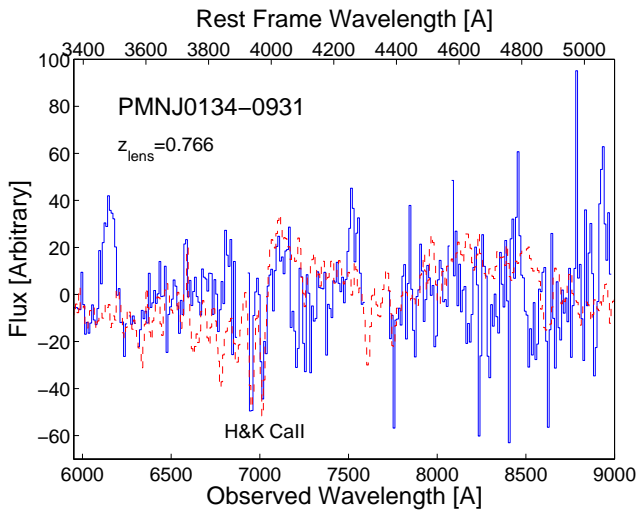}}
\caption {Spectrum of the lensing galaxy
in PMNJ0134$-$0931 (solid line),
after subtraction of a highly smoothed version of itself.
The dashed line shows a galaxy spectral template after 
subtraction of a smoothed version of itself, normalized
to have a similar variance as the spectrum of the lensing galaxy.
\label{PMNJ0134} }
\end{figure}

\begin{figure}
\centerline{\includegraphics[width=8.5cm]{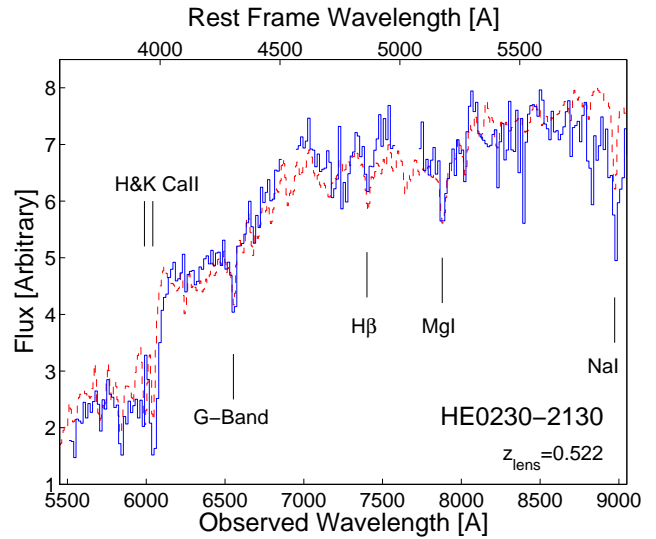}}
\caption {Same as Fig.~\ref{HE0047},
for HE0230$-$2130.
\label{HE0230} }
\end{figure}

\begin{figure}
\centerline{\includegraphics[width=8.5cm]{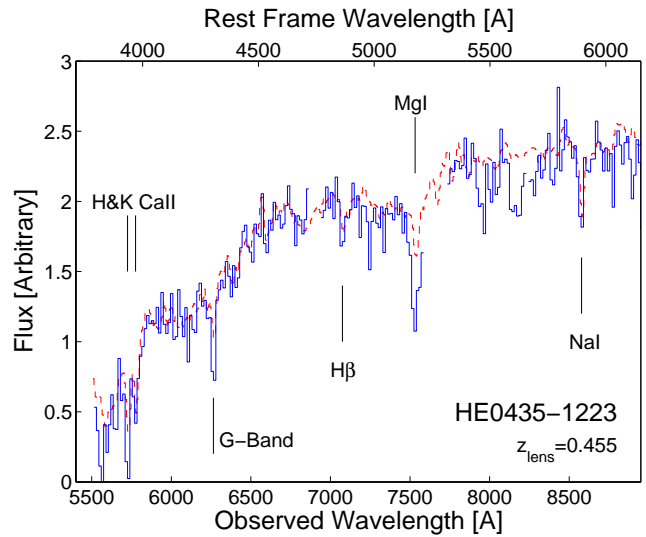}}
\caption {Same as Fig.~\ref{HE0047},
for HE0435$-$1223.
\label{HE0435} }
\end{figure}

\begin{figure}
\centerline{\includegraphics[width=8.5cm]{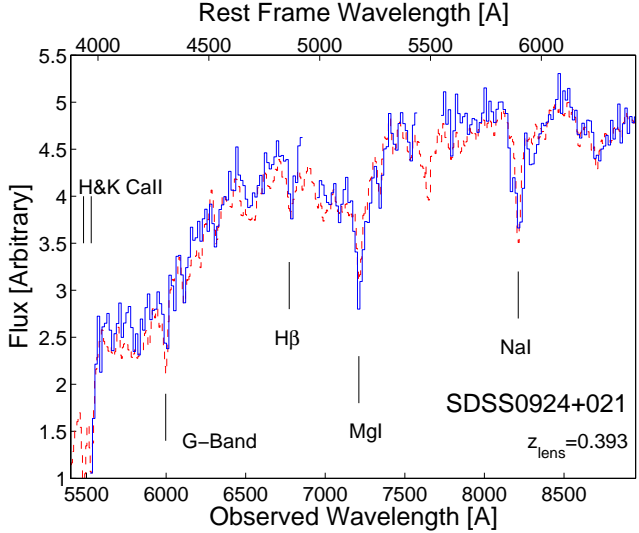}}
\caption {Same as Fig.~\ref{HE0047},
for SDSS0924$+$021.
\label{SDSS0924} }
\end{figure}

\begin{figure}
\centerline{\includegraphics[width=8.5cm]{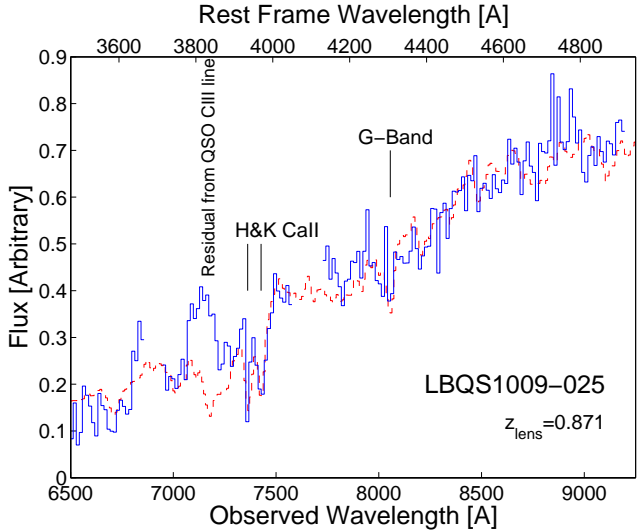}}
\caption {Same as Fig.~\ref{HE0047},
for LBQS1009$-$025.
\label{LBQS1009} }
\end{figure}

\begin{figure}
\centerline{\includegraphics[width=8.5cm]{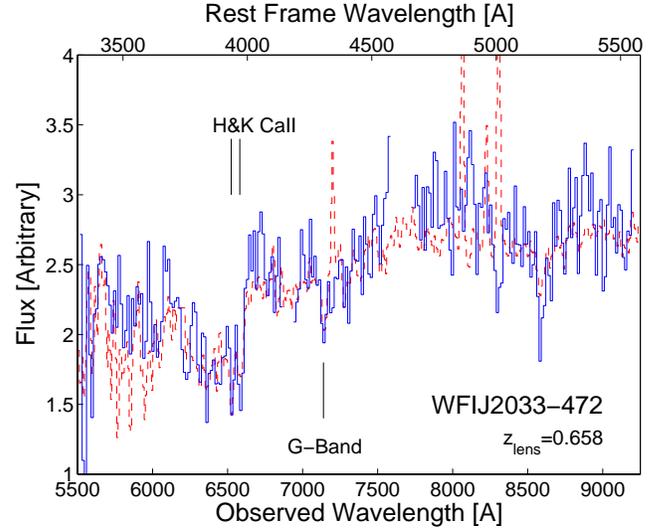}}
\caption {Same as Fig.~\ref{HE0047},
for WFIJ2033$-$472.
\label{WFIJ2033} }
\end{figure}

We trace the spectra by fitting the fluxes and widths of Gaussians models for
each component in the slit as a function of wavelength,
where the relative spatial positions of the Gaussians
were fixed using high-precision astrometry from HST images
of the lenses (Falco et al. 2001 and the CASTLES database\footnote{http://cfa-www.harvard.edu/castles/index.html}).
We extracted a spectrum at the position of the galaxy (contaminated by the quasar light)
and a spectrum of the quasar from a position along the slit 
that is diametrically opposed to, and
as far as possible
from, the galaxy position, but still having sufficient signal-to-noise (S/N)
ratio.
The spectra were
corrected for atmospheric extinction
using
standard extinction curves\footnote{http://www.eso.org/},
and 
flux calibrated using the standard star LTT1788.
A ``cleaned'' galaxy spectrum (i.e., de-contaminated of quasar light)
was obtained by subtracting a scaled version of the 
quasar spectrum from the blended galaxy spectrum.
The proper scaling of each of the quasar spectra was found by visual inspection of
the subtracted spectra of the galaxy,
and by requiring that the quasar emission lines vanish from the decontaminated galaxy spectrum.
The galaxy fraction in the blended quasar+galaxy spectrum was in the
range $~\sim 0.04$ to $\sim 0.9$, with a median of $0.25$.
%
Figures~\ref{HE0047}-\ref{WFIJ2033} show the spectra of the lensing galaxies
for which we succeed in measuring redshifts.
In the other four cases, low S/N in the galaxy spectrum,
excessive contamination from the quasars,
or other peculiarities of the spectrum (see below), prevented us from
measuring secure lens redshifts.

\begin{figure*}
\centerline{\includegraphics[width=18cm]{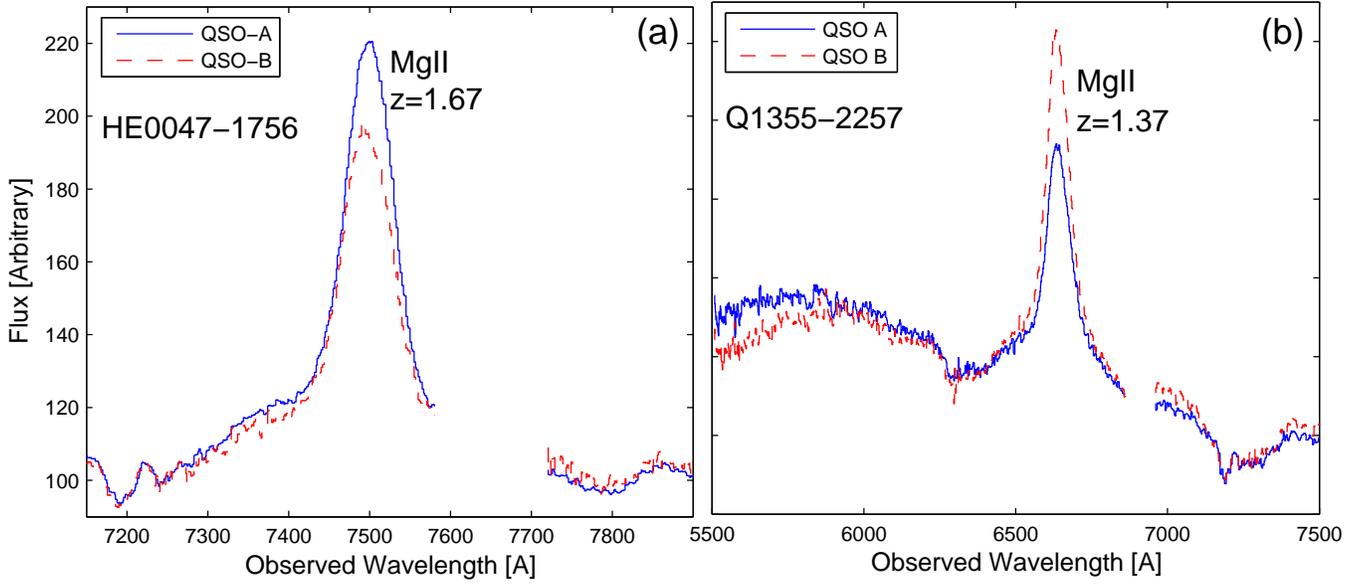}}
\caption {The Mg~{\small II}~$\lambda 2800$~emission
lines of the two lensed quasar images in HE0047$-$1756 (panel a)
and in Q1355$-$2257 (panel b).
After the continua are scaled to match each other,
there are $13\%$ and $60\%$ differences, respectively, in the
emission line fluxes of the two systems.
The spectral differences could be due to microlensing,
millilensing, or intrinsic quasar variability
coupled with the time delay between images.
The spectra of Q1355$-$2257 
diverge from each other
shortward of $\sim5870$~\AA,~which
is the expected wavelength of
the Balmer discontinuity of a $z=0.48$ galaxy (see text).
\label{QSOs_HE0047_Q1355} }
\end{figure*}

Our extraction process is not free of complications. Lensed quasars show
differences among lensed images in both
the continua and the emission lines (e.g., Richards et al. 2004; Keeton et al. 2005).
For example, in Fig.~\ref{QSOs_HE0047_Q1355}-a
we show the Mg~{\small II}~$\lambda 2800$~emission line of the two images of HE0047$-$1756.
After matching the continua of the two quasar images to the same level,
there is a $13\%$ difference in the flux of the Mg~{\small II} lines.
This difference is not due to the galaxy. The total emission
from the galaxy, most of which lies in
our galaxy aperture rather than the quasar apertures, is
only $10\%$ that of the quasars, and the spectrum of the galaxy
has no feature near the Mg~{\small II}~$\lambda 2800$~emission line
that could produce an apparent difference between the line and continuum
of the quasar.
Figure~\ref{QSOs_HE0047_Q1355}-b shows a second example, the
spectra of both quasar images of Q1355$-$2257 in the vicinity of the
Mg~{\small II}~$\lambda 2800$~emission line. The line equivalent widths
differ by about $60\%$.
Such differences in quasar spectra can be the result of
microlensing
that is preferentially magnifying or demagnifying the continuum
or the emission lines of the images
(e.g., Schechter \& Wambsganss 2002; Keeton et al. 2005),
millilensing (e.g., Abajas et al. 2002; Lewis \& Ibata 2004; Keeton et al. 2005),
or intrinsic quasar variability (e.g., Kaspi et al. 2000; 2005) combined
with the time delay between the images (about $50$~days and $40$~days
for HE0047$-$1756 and Q1355$-$2257, respectively; Witt, Mao, \& Keeton 2000).
The spectral differences between the two images
could affect the galaxy spectrum that is obtained by subtracting
the spectrum of quasar A from 
the lensed spectrum at the galaxy position,
which is located between images A and B (see Fig.~\ref{AllFindingCharts}).
An example of this effect is seen in the spectrum of LBQS1009$-$025 (Fig.~\ref{LBQS1009}),
in which a residual of the quasar's C~{\small III]}~$\lambda1909$~emission line is left in the
``cleaned'' galaxy spectrum.
Note that, in this case, the lensing galaxy
is located on top of image B (see Fig.~\ref{AllFindingCharts})
and hence there is no choice but to use image-A for the subtraction.


The redshifts of each galaxy were
obtained by cross correlating the spectrum with template local galaxy spectra
from Kinney et al. (1996).
For each galaxy we tried template spectral types 
from elliptical to Sc, or ``mixes'' of varying weights between two 
adjacent types. 
To each template we first applied the redshift and then the
Galactic extinction (Schlegel, Finkbeiner, \& Davis 1998)
toward the lens.
In one case, PMNJ0134$-$0931, the lensing galaxy was not detected
directly
(by the procedure described above), but several absorption
lines are visible in the quasar spectrum.
In this case, we subtracted from the quasar spectrum a smoothed
version of itself, and compared the resulting absorption line spectrum with
a similarly normalized galaxy template.

Figures~\ref{HE0047}-\ref{WFIJ2033} show, along with the spectra of the lensing galaxies,
also the best matching galaxy templates.
%
%
The measured redshifts of the lensing galaxies,
the Galactic extinctions (Schlegel et al. 1998),
and the quasar redshifts, are summarized in Table~\ref{Table-z}.
The redshifts of two of the lensing galaxies 
have been previously measured.
The lens redshift of PMNJ0134$-$0931 was previously measured to be $z=0.7645$,
by  Hall et al. (2002, based on the H\&K~Ca~{\small II} absorption)
and by Kanekar \& Briggs (2003, based on $21$~cm HI absorption).
During the course of our VLT program, the lensing galaxy redshift in
HE0435$-$1223 was measured by
Morgan et al. (2005) to be $z=0.4546\pm0.0002$.
In both cases, we confirm the measured redshifts.

Although the continuum slopes and absorption features
of three of the lens galaxies
are best fit by spiral, rather than elliptical or S0 templates, 
in most cases there is evidence that
the lens galaxies are nevertheless early types,
which at high redshifts may have bluer colors than those of local
ellipticals, probably due to some post-starburst signatures.
Indeed, the [O~{\small III}] emissions lines in the spiral templates
are always absent in the lens galaxies, indicating a lack of
ongoing star formation. In two cases, the intrinsically blue color
may be compounded by some mild distortion of the continuum slope
as a result of differential atmospheric refraction affecting our spectra, which
were obtained
at a non-parallactic angle.
The observational and evolutionary effects are both discussed
on a case by case basis in \S\ref{Notes}.

\begin{deluxetable}{llcc}
\tablecolumns{4}
\tablewidth{0pt}
\tablecaption{Lensing galaxies - measured redshifts}
\tablehead{
\colhead{Object Name} &
\colhead{$z_{lens}$} &
\colhead{Galactic $E_{B-V}$} &
\colhead{$z_{QSO}$}
}
\startdata
HE0047$-$1756   & $0.408$                 & $0.022$ & $1.67 $ \\      
PMNJ0134$-$0931 & $0.766$\tablenotemark{a}& $0.031$ & $2.216$ \\
HE0230$-$2130   & $0.522$                 & $0.022$ & $2.162$ \\      
HE0435$-$1223   & $0.455$\tablenotemark{b}& $0.068$ & $1.689$ \\      
SDSS0924$+$021  & $0.393$                 & $0.055$ & $1.524$ \\      
BRI0952$-$0115  &                         & $0.063$ & $4.426$ \\
LBQS1009$-$025  & $0.871$                 & $0.034$ & $2.74 $ \\      
Q1017$-$207     & $1.088?$                & $0.046$ & $2.545$ \\
Q1355$-$2257    & $0.48?$                 & $0.072$ & $1.373$ \\
PMNJ1632$-$003  & $1.165?$                & $0.098$ & $3.424$ \\
WFIJ2033$-$472  & $0.658$                 & $0.047$ & $1.66 $ \\      
\enddata
\tablenotetext{a}{Redshift previously measured by Hall et al. (2002), and Kanekar \& Briggs (2003).}
\tablenotetext{b}{Redshift previously measured by Morgan et al. (2005)}.
\tablecomments{The uncertainty in redshift for all lenses is $\sim0.001$,
and is dominated by the error in
the cross correlation, with a smaller contribution from the wavelength calibration uncertainty.
The cross-correlation uncertainty was estimated based on a $\Delta\chi^{2}$ test.}
\label{Table-z}
\end{deluxetable}

\begin{figure*}
\centerline{\includegraphics[width=18cm]{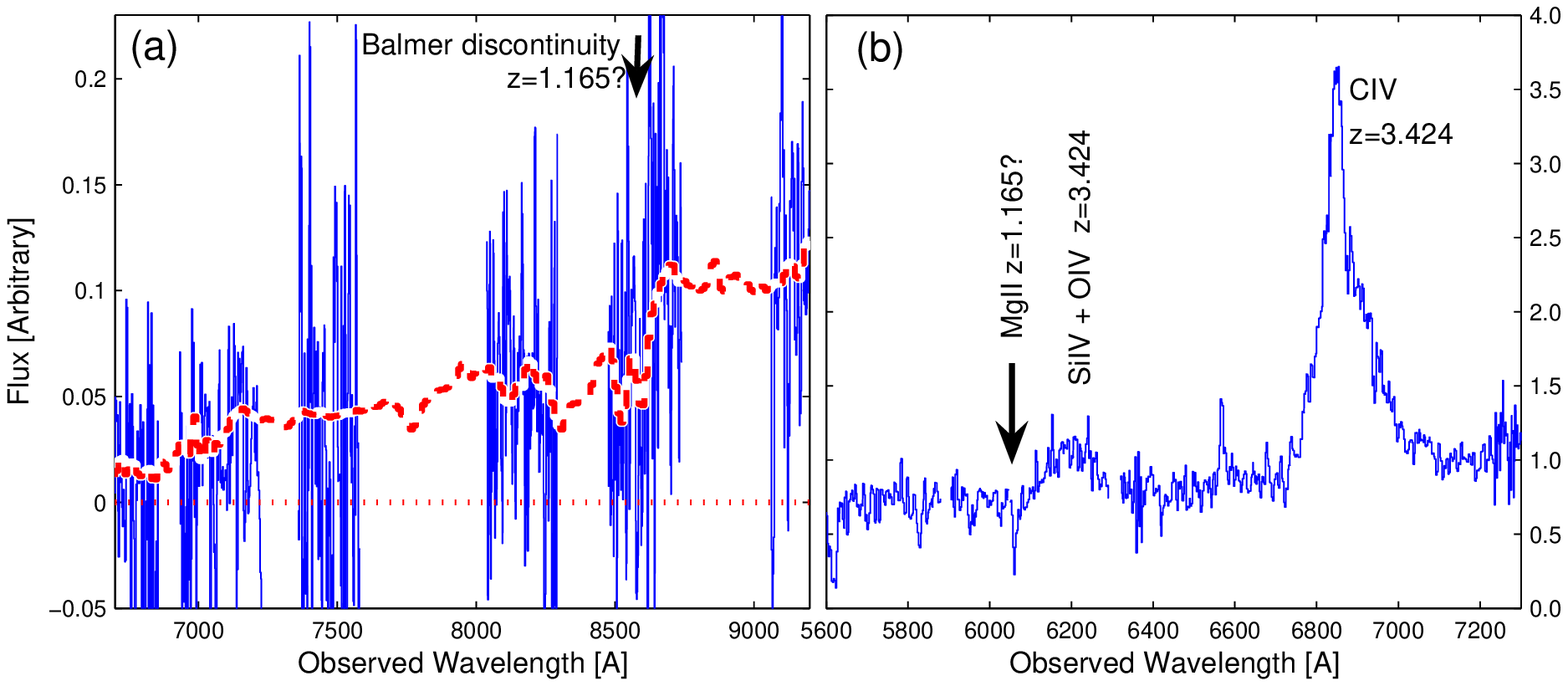}}
\caption {Panel a: Lens galaxy spectrum for PMNJ1632$-$003.
Gaps in the spectra correspond to regions with very low
S/N, due to atmospheric emissions,
which we do not display for the sake of clarity.
The dashed line is an elliptical galaxy template, redshifted to
$z=1.165$
Panel b: Spectrum of image-A of the lensed quasar
PMNJ1632$-$003.
The absorption at $6060$~\AA~could be 
Mg~{\small II}~$\lambda 2800$~\AA, at the redshift suggested by the 
break seen in the galaxy spectrum in panel (a).
\label{PMNJ1632_Gal_QSO} }
\end{figure*}

\section{Notes on individual objects}
\label{Notes}

{\bf HE0047$-$1756}:
A doubly imaged lensed quasar discovered by Wisotzki et al. (2004),
with an image separation of $1\farcs44$.
The continuum and absorption features of the lens galaxy spectrum, 
shown in Fig.~\ref{HE0047},
are best matched
by an Sb$+$Sc template (with a mix ratio of $9:1$), at redshift $z=0.408$.
We note, however, that the emission lines in the template are absent
from the lens galaxy spectrum. To investigate the source of this disparity,
we have first examined whether the blue color of the lens galaxy 
could be the result of light loss through the slit,
as a result of the 
differential atmospheric refraction\footnote{http://www.eso.org/gen-fac/pubs/astclim/lasilla/diffrefr.html}
(Fillipenko 1982) and the non-parallactic slit angle that we used. We have estimated
at $\lesssim 15\%$ the possible change in spectral slope due to such light loss,
for the particular slit angle, declination, observed hour angles 
and seeing widths appropriate for each exposure of this object.
The corresponding difference in spectral slopes between the best-fit template and 
that of an elliptical galaxy is $\sim 70\%$. Furthermore, since the slit-centering  acquisition 
images were obtained in the $z$ band, we would expect that atmospheric refraction
would lead to blue, rather than red, light loss, and hence to a reddening of the continuum.
The blue color is therefore unlikely
to be an observational defect. 

A rough test of whether the 
mass of the lens is consistent with that of a normal spiral galaxy can be obtained
by evaluating the velocity dispersion of a singular isothermal sphere (SIS)
that would produce the observed image separation, given the source and lens
redshifts,
\begin{equation}
\sigma_{\rm SIS}^{2}=\frac{c^2}{4\pi}\frac{D_{s}}{D_{ls}}\frac{\theta}{2},
\end{equation}
where $\theta$ is the image separation.
For the case at hand, $\sigma_{\rm SIS}\approx190$~km~s$^{-1}$, which is typical
of early-types, but somewhat large for spirals. Along with the absence of
indications of ongoing star formation, this argues that the lens galaxy
is of early type. Its relatively blue color may then be the result of
post-starburst signatures that are still present at the $\sim 4$~Gyr lookback time.      

{\bf PMNJ0134$-$0931}:
A highly reddened lensed quasar with a complex radio morphology, discovered by
Winn et al. (2002a).
The quasar is at redshift $z=2.216$. The two brightest images
are separated by $0\farcs15$ (see Fig.~\ref{AllFindingCharts}).
The lensing galaxy redshift was 
measured by Hall et al. (2002)
and by Kanekar \& Briggs (2003)
and found to be $z=0.7645$.
This is consistent with our result, $z=0.766\pm0.001$,
based on the spectrum shown in Fig.~\ref{PMNJ0134}.

{\bf HE0230$-$2130}:
Discovered by Wisotzki et al. (1999), this is a quadruply
lensed quasar with redshift $z=2.162$ and
a maximum image separation of $2\farcs1$. 
The lensing galaxy spectrum, shown in Fig.~\ref{HE0230},
is well matched by an equal mix of a S0 and Sa templates,
with a redshift of $z=0.522$. Following the same test
as in the case of HE0047$-$1756, above, we find that 
the non-parallactic slit angle could lead to a change in slope
of $\lesssim 30\%$, and as before, this would likely be in the
sense of making the object redder. Between the S0 and Sa 
templates, the slope in the observed frame 6000-8500~\AA~ range changes
by $\sim 70\%$. Thus, observational effects could possibly play a
partial role
in the blue color of the lens galaxy. In terms of galaxy mass,
the derived SIS velocity dispersion in this case is
$\sigma_{\rm SIS}\approx240$~km~s$^{-1}$, which would be highly unusual
for a spiral galaxy. We conclude that the lens galaxy is of early type
with a blue color probably due to its being observed at a lookback time of 
5~Gyr.
      
{\bf HE0435$-$1223}:
Discovered by Wisotzki et al. (2002),
with a quasar redshift of $1.689$. This is a quadruply lensed system with
a maximum image separation of $2\farcs6$. 
The lens redshift was recently measured at $z=0.4546$ (Morgan et al. 2005).
We confirm this result and find that the galaxy spectrum,
presented at Fig.~\ref{HE0435},
is well matched by an S0 template.

{\bf SDSS0924$+$021}:
Discovered as part of the Sloan Digital Sky Survey by Inada et al. (2003).
This is a quadruply lensed quasar with a maximum image
separation of $1\farcs8$, in which the merging pair   
(images A \& D in Fig.~\ref{AllFindingCharts})
has a broad-band flux ratio of $10$ --
the most extreme known case of a
``flux ratio anomaly''
(see Keeton et al. 2005, for implications).
The lensing galaxy spectrum, shown in Fig.~\ref{SDSS0924},
is well matched by an E galaxy template,
with a redshift of $0.393$.

{\bf BRI0952$-$0115}:
A doubly imaged quasar, with $1\farcs0$ separation,
discovered by McMahon \& Irwin (1992).
At $z=4.426$ (the redshift measurement 
was refined by Storrie-Lombardi et al. 1996),
this is the highest redshift lensed quasar known.
The ``fundamental-plane redshift'' of the
lensing galaxy is $z=0.41\pm0.05$ (Kochanek et al. 2000).
Storrie-Lombardi et al. (1996) identified several metal
absorption systems, at redshifts of
$1.993$, $3.294$, $3.475$, $3.719$, and $4.024$ in the spectrum of the
lensed quasar, one or more of which could contribute to the lensing.
We note, however, that if these absorbers indeed contribute to
the lensing then the fundamental-plane redshift
may be biased toward lower redshifts.
Even with over $3$~hours of exposure time, the signal from
the galaxy is 
too low to securely identify
any features in the spectrum.
%
%

%

{\bf LBQS1009$-$025}:
A doubly imaged quasar, with $1\farcs54$ image separation,
discovered by Hewett et al. (1994).
The quasar is at a redshift of $2.74$.
The lensing galaxy spectrum, shown in Fig.~\ref{LBQS1009},
is matched by an E galaxy template
with a redshift of $0.871$, very close to the
fundamental-plane redshift ($0.88$) found by
Kochanek et al. (2000).

\begin{figure}
\centerline{\includegraphics[width=8.5cm]{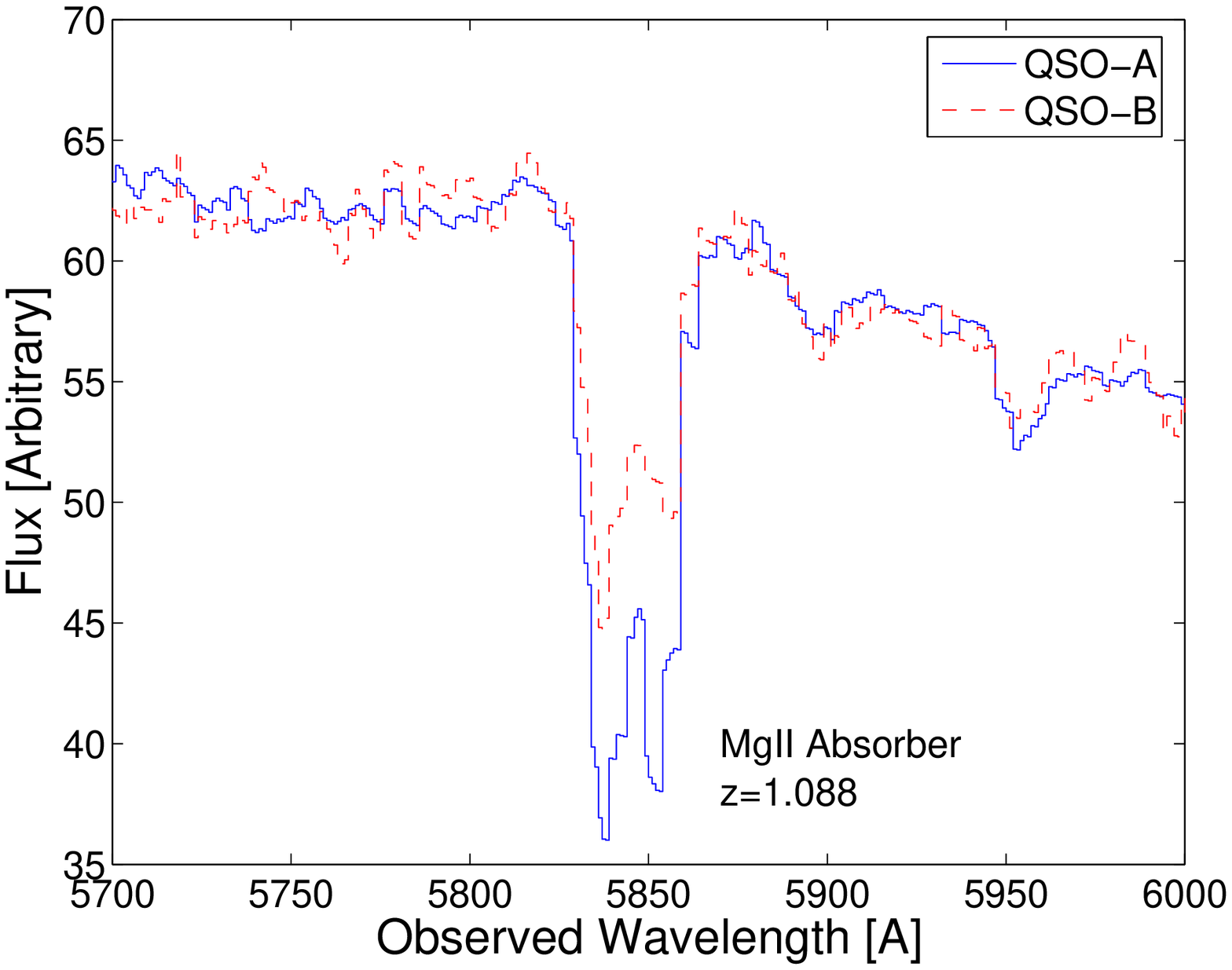}}
\caption{Section of the spectra of the two lensed images
of Q1017$-$207, after the continua are scaled to match each other.
A Mg~{\small II} absorber at $z=1.088$ is detected in the spectra of both
images, and given the expected blending,
we suspect that the absorber is present in both images (see text).
\label{Q1017_MgII_Abs} }
\end{figure}

{\bf Q1017$-$207}:
A doubly imaged quasar, with $0\farcs85$ image separation,
discovered by Claeskens et al. (1996).
The quasar is at a redshift of $2.545$ and
the fundamental-plane redshift of the
lensing galaxy is $z=0.78\pm0.07$ (Kochanek et al. 2000).
Due to limited S/N
(only $1.6$~hr, of a requested $4.0$~hr exposure time,
were actually obtained)
we fail to
identify the type or redshift of the lens galaxy.
Claeskens et al. (1996) detected a Mg~{\small II}~$\lambda 2800$ absorber,
with $z=1.085$,
in the combined spectra of the images, and suggested
that it could be due to the lensing galaxy.
Surdej et al. (1997) obtained HST Faint Object Spectrograph (FOS)
spectra of both images and detected the Mg~{\small II}
absorber only in the spectrum of image A.
In our VLT spectra,
the Mg~{\small II} absorber is apparent in
both images, and we refine its redshift to  $z=1.088\pm0.001$ (Fig.~\ref{Q1017_MgII_Abs}).
The absorption equivalent widths are
$9.4$~\AA~and $5.5$~\AA~(observed frame)
for image A and B, respectively.
After correcting for the expected contamination by image A,
we estimate that the Mg~{\small II}
absorption equivalent width in image B is
$4.2$~\AA~(observed frame).
The correction was made
by calculating the ratio between the
blended flux of image B
(in the $5700-5800$~\AA~range), to the expected
flux of image A at the same cross-dispersion position,
based on the point-spread-function wing of image A
on the side opposite to image B.
%
By inspecting the HST-FOS spectra (Fig.~2 in Surdej et al. 1997),
we estimate
that the S/N ratio of the VLT spectrum is $10-20$
times higher than that of the HST/FOS spectra.
This may explain the absence of
the Mg~{\small II} absorber in image B in the HST/FOS spectra.
If, as we suspect, the Mg~{\small II}
absorber is indeed present in the image B spectrum,
the physical separation at the redshift of the absorber,
about $7$~kpc,
suggests that the absorber is a massive galaxy capable
of being the lens.

{\bf Q1355$-$2257}:
A doubly imaged quasar, with $1\farcs23$ image separation,
discovered by Morgan et al. (2003).
The quasar is at a redshift of $1.373$.
Based on the flux of the galaxy and 
the Faber-Jackson (1976) relation,
Morgan et al. (2003) estimated that the lens galaxy lies
in the redshift range $0.4<z<0.6$,
and suggested that the 
Mg~{\small II}~$\lambda\lambda2796,2803$~\AA~absorption
feature at $z=0.48$ identified in their
HST/FOS spectra of the system, could be
associated with the lens galaxy.
We note that our spectrum of quasar B,
which would capture the most flux from the lens galaxy, shows
an excess of emission relative to quasar A at wavelengths longward
of $\sim5870$~\AA~(see Fig.~\ref{QSOs_HE0047_Q1355}).
This roughly corresponds to the wavelength at which a $z=0.48$
galaxy would have its $4000$~\AA~break.

{\bf PMNJ1632$-$003}:
A doubly imaged quasar, with $1\farcs47$ image separation,
discovered in radio frequencies by Winn et al. (2002b).
The quasar is at a redshift of $3.424$,
and is the only quasar lensed by a galaxy
for which the demagnified central image
has been securely detected (Winn et al. 2004).
The existence of the central image sets a
lower limit on the
galaxy's surface density at the location of the central image.
The spectrum of the lensing galaxy is noisy and we cannot
confidently identify the redshift.
However, as seen in Fig.~\ref{PMNJ1632_Gal_QSO}-a,
we detect a jump in the spectrum at $\sim8600\pm40$~\AA.
If attributed to the Balmer discontinuity, then
an absorption line at $6060$~\AA~in the QSO spectrum
(Fig.~\ref{PMNJ1632_Gal_QSO}-b) may be
Mg~{\small II}~$\lambda2800$~\AA,
and the galaxy redshift would be  $1.165$.

{\bf WFIJ2033$-$472}:
A quadruply lensed quasar, discovered by Morgan et al. (2004).
The quasar redshift is $1.66$ and the maximum image separation
is $2\farcs5$.  
The lensing galaxy spectrum, shown in Fig.~\ref{WFIJ2033},
is matched by an Sb and Sc galaxy template
(mix ratio of $4:1$)
with a redshift of $0.658$.
As in the cases of HE0047$-$1756 and HE0230$-$2130, above, 
the non-parallactic slit angle could lead to a change in slope
of $\lesssim 30\%$, and as before, this would likely be in the
sense of making the object redder. Between the S0 and Sb/Sc 
templates, the slope in the 6000-8500~\AA~ range changes
by $\sim 75\%$ and therefore differential atmospheric refraction
 is unlikely  to be the main cause of 
the blue color of the lens galaxy. 
The derived SIS velocity dispersion in this case is
$\sigma_{\rm SIS}=300$~km~s$^{-1}$, corresponding to a massive elliptical
or a galaxy group. As in the previous two cases, the spiral-galaxy-like
slope and absorption features in the lens galaxy spectrum 
most probably do not indicate a spiral galaxy lens.


To summarize,
we have obtained low resolution VLT/FORS2 optical spectra of
$11$ gravitationally lensed quasar systems,
and we have measured the redshifts of seven
of the lensing galaxies.
In three cases, the best spectral fits correspond to 
local spiral galaxy templates in terms of continuum shape
and absorption feature depths. However, in all three cases
we have argued that the lens galaxies are actually 
of early-type, based on the absence of emission lines,
and on the large image separations that they produce, which
are characteristic of massive ellipticals. In each case,
we have also argued against a dominant observational source
for the anomalous colors, and concluded that these
early-type galaxies, as viewed at lookback times of $\sim 5$~Gyr,
are intrinsically blue. Their observed spectra likely carry 
the traces of previous episodes of star formation. 
Our fairly high success rate in obtaining spectroscopic redshifts,
shows that full redshift information can be gathered for most lensed
quasar systems. As discussed in \S1,
this information will be useful for a host of cosmological applications.

\acknowledgments
We thank Dovi Poznanski, Orly Gnat, and Shai Kaspi,
for valuable discussions.
This work was supported by a grant from the
German Israeli Foundation for Scientific Research and Development.

\end{document}